\documentstyle[aps,prl,preprint]{revtex}

\newcommand{\half}{{\scriptstyle{\frac{1}{2}}}}

\newcommand{\kk}{{\bf k}}
\newcommand{\pp}{{\bf p}}
\newcommand{\qq}{{\bf q}}

\newcommand{\BE}{\begin{equation}}
\newcommand{\EE}{\end{equation}}
\newcommand{\BA}{\begin{eqnarray}}
\newcommand{\EA}{\end{eqnarray}}

\begin{document}

\title{Light Higgs bosons from a strongly interacting Higgs sector}

\author{Fabio Siringo}
\address{Dipartimento di Fisica e Astronomia, 
Universit\`a di Catania,\\
Corso Italia 57, I 95129 Catania, Italy}
\date{\today}
\maketitle
\begin{abstract}
The mass and the decay width of a Higgs boson in the minimal
standard model are evaluated by a variational method in the
limit of strong self-coupling interaction. The non-perturbative
technique provides an interpolation scheme between strong-coupling
regime and weak-coupling limit where the standard perturbative
results are recovered. In the strong-coupling limit the
physical mass and the decay width of the Higgs boson are found
to be very small as a consequence of mass renormalization.
Thus it is argued that the eventual detection of a light Higgs
boson would not rule out the existence of a strongly interacting 
Higgs sector.
\end{abstract}
The impressive success of the Standard Model (SM) has enforced
the common believing
that the Higgs boson will be soon detected by
the new generation of accelerators\cite{altarelli}.
In fact there are two unknown parameters in the SM that wait for their 
experimental determination: the mass $m$ of the Higgs boson and
the strength of its
self-coupling interaction $\lambda$. This last one determines
the bare Higgs mass
$m_0^2=\lambda v^2/3$ where $v$ is the vacuum expectation value
for the scalar field
which is fixed by the known strength of weak interactions.
Thus at tree level perturbation
theory predicts a light Higgs mass $m\approx m_0$ if the coupling
$\lambda$ is small enough.
Conversely, in the strong coupling limit, perturbation theory breaks down and
there is no simple
relation between $m$ and $\lambda$.
A light weakly interacting Higgs boson has been
strongly desired, mainly because perturbation theory would be reliable,
and the Higgs boson would be detectable at a reasonable energy threshold. 
However, if nature had chosen for a strongly interacting boson, the physics
would be richer and more interesting. Actually, the physics of such a strongly 
interacting Higgs boson has been explored in the last twenty years, 
and interesting proposals have been discussed ranging from the existence 
of bound states\cite{suzuki,5inrupp,grifols,rupp,DDh,clua,DDg} 
to unconventional descriptions of the 
symmetry breaking mechanism\cite{consoli}.

During the last years the possibility of a strongly interacting Higgs boson
has been rejected for two main reasons: i) A large $\lambda$ is
believed to imply a large mass, in contrast with the recent phenomenological
evidence\cite{altarelli} for a light $m\approx 100-200$ GeV; ii) For a strongly 
interacting Higgs boson the decay width $\Gamma$ has been predicted 
to be very large\cite{herrero,1/N} compared to the mass, 
and such very large resonance could hardly be regarded
as a true particle. In this letter we point out that both the statements i) 
and ii) have a perturbative nature and cannot
be trusted in the strong coupling limit.
At tree level $m$ and $\Gamma$ are small if the coupling $\lambda$ is
small, which is consistent in the framework of perturbation theory. 
However if $\lambda$ is very large any perturbative argument breaks down
and fails to predict what $m$ and $\Gamma$ are.
In fact, by use of a variational method we show that
both $m$ and $\Gamma$ are small in the strong coupling limit.

The existence of a saturation of $m$ at strong coupling has been
shown by several non-perturbative techniques
as $1/N$ expansions\cite{1/N}, variational methods\cite{varhiggs}
and Bethe-Salpeter equation\cite{rupp2}.
We have shown that a further increase of the coupling strength yields
a decrease of the mass\cite{varhiggs},
and this has also been confirmed by recent
Bethe-Salpeter calculations\cite{rupp2}.
The physical reason is very simple:
at tree level $m$ is proportional to $\lambda$; however the interaction
renormalizes the mass, since the attractive self-coupling reduces
the energy of a free boson. At some stage this reduction overcomes
the tree level increase, and the renormalized mass decreases for some
very large self coupling. As a result a light Higgs boson could be a very
strongly interacting particle whose ground state could even be a
Higgs-Higgs bound state.

A light self-interacting Higgs boson would not make any sense as a free
particle if its decay width $\Gamma$ would be so large and increasing
with $\lambda$ as found by $1/N$ expansion calculations\cite{1/N}. 
However in the
real world the goldstone bosons of the $O(N)$ model do not play any
physical role, while the Higgs sector is coupled with the gauge bosons
through a quite weak interaction which does not increase with $\lambda$.
As could be expected, we show that for very large couplings and a
reasonable choice of the cut-off, a light Higgs boson would be
characterized by a very small decay width: thus the experimantal
knowledge of $m$ and $\Gamma$ would not say the last word on the
strength of the self-interaction. The eventual detection of a light
Higgs with a narrow decay width would be consistent with both a
perturbative weakly interacting and a non-perturbative strongly
interacting theory.

In order to deal with the non-perturbative limit we use a
variational method in the Hamiltonian
formalism\cite{schiff,stevenson,DDe,DDh,DDg}.
The method has the advantage of yielding the known perturbative results
in the weak-coupling limit\cite{DDe,DDg}
(e.g. masses, decay widths and binding energies), while it can be safely
extended to the non-perturbative strong coupling regime.
The results achieved by such method have not been appreciated in the past
since the variational equations have been usually approximated by
perturbative methods\cite{perturb} thus spoiling their most important
advantages. In the framework of a study on bound states we have recently
shown\cite{varhiggs} that the variational equations can be decoupled
exactly, giving important consequences on mass renormalization. In this
letter we show that the same method can be used for decoupling the
variational equations arising from a more complete trial state,
describing a Higgs field $h$ which interacts with a neutral
gauge vector field $Z^\mu$:
\BE
\vert\Psi\rangle=\vert h\rangle+\vert hh\rangle+\vert hhh\rangle
+\vert ZZ\rangle
\label{trial}
\EE
where
\BE
\vert h\rangle=A  a^\dagger_0 \vert 0 \rangle, 
\label{h}
\EE
\BE
\vert hh\rangle=\int d^3 p \, 
B(\pp) a^\dagger_\pp a^\dagger_{-\pp}\vert 0
\rangle,
\label{hh}
\EE
\BE
\vert hhh\rangle=
\int d^3 p \, d^3 q \, d^3 k \, 
G(\pp,\qq,\kk) a^\dagger_\pp a^\dagger_\qq
a^\dagger_\kk \vert 0\rangle \delta^3 (\pp+\qq+\kk),
\label{hhh}
\EE
\BE
\vert ZZ\rangle=
\sum_{\sigma\sigma^\prime}
\int d^3 p \, 
C_{\sigma\sigma^\prime}(\pp) 
b^\dagger_{\pp\sigma} b^\dagger_{-\pp\sigma^\prime}
\vert 0 \rangle.
\label{ZZ}
\EE
Here $a^\dagger_\pp$ is the creation operator 
for a Higgs particle of momentum $\pp$ and mass $m$, 
$b^\dagger_{\pp\sigma}$ is the creation operator  
for a neutral vector boson $Z^0$ of momentum $\pp$,
polarization $\sigma$ and mass $M$, and $\vert 0\rangle$
is the vacuum annihilated by the corresponding
annihilation operators. The coefficients $A,B,C,G$
can be determined from the variational principle
\BE
\delta\langle\Psi\vert: \hat H-E:\vert\Psi\rangle=0.
\label{variation}
\EE
All the required terms of the Hamiltonian $\hat H$
can be canonically derived from the SM Lagrangian
density
\BA
{\cal{L}}=
-\half\partial_\mu h\partial^\mu h
-\half m_0^2 h^2&-&{1\over{3!}}\lambda v h^3-{1\over{4!}}\lambda h^4
-{1\over 4} F_{\mu\nu}F^{\mu\nu}-\nonumber\\
&-&\half M^2 Z_\mu Z^\mu
-{{M^2}\over{v}} Z_\mu Z^\mu h-\half\left({M\over v}\right)^2
Z_\mu Z^\mu h^2. 
\label{lagrangian}
\EA
This is the Lagrangian of a $U(1)$ Higgs model
(scalar electrodynamics) which is equivalent
to the full SM Lagrangian as far as we only consider the trial
state (\ref{trial}). The variational principle (\ref{variation})
yields four coupled integral equations (eigenvalue equations)
for the coefficients
$A,B,C,G$. The full equations have been reported in Ref.\cite{DDg}.
They can be considerably simplified by taking advantage of the
symmetry properties of the bosons: without any loss of generality
the functions $B$ and $G$ may be taken to be even under spatial inversion,
the function $G$ may be assumed invariant under any permutation of its
arguments, and we may take
$C_{\sigma\sigma^\prime}(\pp)=C_{\sigma^\prime \sigma} (-\pp)$.
Moreover, up to a vacuum renormalization, we may assume
$G(0,\pp,-\pp)=0$ in the trial state. An exact decoupling
can be easily achieved by the method of Ref.\cite{varhiggs}, thus avoiding
any further approximation. The full details will be published elsewhere.
Here we discuss the results in the two special cases $C=0$ and $G=0$.

For $C=0$ there is no decay and the trial state (\ref{trial}) is an
improvement over the $\vert hh\rangle+\vert hhh\rangle$ variational
ansatz of Ref.\cite{varhiggs}. Here we have one extra equation
arising from the variation with respect to $A$ in Eq.(\ref{variation}).
However the extra coefficient $A$ is a constant which can be easily
eliminated yielding two coupled integral equations. We regularize the
logarithmically divergent integrals with an energy cut-off
$\omega_p=\sqrt{\pp^2+m^2}<\Lambda$. Neglecting terms of order
${\cal{O}}(\Lambda^{-2})$, the method of Ref.\cite{varhiggs} allows an
exact decoupling of the integral equations yielding
\BE
(2\omega_k-E) B(\kk)=-\int \! d^3p \> {\cal {K}}(\kk,\pp,-\kk-\pp) B(\pp) ,
\label{eqint}
\EE
where the kernel ${\cal{K}}$ is defined as 
\BA
{\cal {K}}(\kk,\pp,\qq)={1\over{64\pi^3\omega_k}}\left({{2m_0^2+m^2}\over
{v^2}}\right)\times\qquad\qquad\qquad\qquad\qquad\qquad
\qquad\qquad\qquad\qquad&&\nonumber\\
\times\left\{\left({{2m^2_0-2m^2}\over{2m_0^2+m^2}}\right)
{1\over{\omega_p}}- {2\over{\omega_p\omega_q}}
{{m^2+2m_0^2+\omega_p(E-2\omega_p)}\over
{\left[\omega_k+\omega_p+\omega_q+{{m_0^2-m^2}\over 2}
\left({1\over{\omega_k}}+{1\over{\omega_p}}+{1\over{\omega_q}}\right)
-E\right]}}\right\}&& .
\label{kernel}
\EA
This differs form the ${\vert h\rangle+\vert hh\rangle}$ calculation
of Ref.\cite{varhiggs} for a decrease of the numerical coefficient
of the first (repulsive) term inside the brackets.
In Eq.(\ref{eqint}) 
a self-consistency condition has been imposed in order
to fix the lower bound $E_0$ of the continuous spectrum
of two-particle scattering states. Imposing
$E_0=2m$ yields the mass renormalization condition
\BE
m^2=m^2_0\left[{{1-2J(0)}\over{1+J(0)}}\right]
\label{self1}
\EE
where
\BE
J(0)={{\lambda}\over{32\pi^2}}\int_1^{\Lambda/m}
{{\sqrt{x^2-1}}\over{x^2-\alpha x
-\beta}}dx,
\label{self2}
\EE
$\alpha=(3-m^2_0/m^2)/4$ and $\beta=(1-m^2_0/m^2)/2$.
These conditions ensure that the integral equation (\ref{eqint})
always admits the free-wave solution $E=2m$ as the lower bound
of the continuos spectrum.  
The numerical solution of the coupled equations (\ref{self1}),
(\ref{self2}) is reported in Fig.1 for a large cut-off
$\Lambda=14$ TeV. The perturbative approximation $m\approx m_0$
breaks down for $m_0> 0.3$ TeV. Moreover, in the strong coupling
limit, we find a light $m<100$ GeV for any $m_0> 1.9$ TeV. 
Thus a physical mass
$m\approx 100$ GeV could result from very small or
very large couplings. The strong coupling case is
characterized by the presence of bound state solutions, i.e.
two-particle solutions of Eq.(\ref{eqint}) with $E<2m$. In
Fig.2 the binding energy is reported and compared to the
prediction of the $\vert hh\rangle+\vert hhh\rangle$ ansatz.
In the present calculation, the presence of the extra term
$\vert h\rangle$ represents an improving of the trial state,
and causes a decrease of the binding energy
as it should be expected for any variational calculation.

In order to study the decay width we must restore $C\not=0$
in the trial state (\ref{trial}). Here we prefer to
discuss the $G=0$ case for brevity. For the
$\vert h\rangle +\vert hh\rangle+\vert ZZ\rangle$ state
the eigenvalue equations can be easily decoupled
yielding
\BE
C_{\sigma\sigma^\prime}(\pp)=\left[e^\mu (\pp\sigma)
e_\mu (-\pp\sigma^\prime)\right]^*
\left[ \delta^3(\pp-\pp_0)-\Delta(E) f(\pp,E)\rho(E)\right]
\label{scattered}
\EE
\BE
\Delta(E) =\sum_{\sigma\sigma^\prime}\int {{d^3 p}\over{\Omega_p}}
e^\mu (\pp\sigma)e_\mu (-\pp\sigma^\prime)
C_{\sigma\sigma^\prime} (\pp)
\label{delta}
\EE
\BE
f(\pp,E)=\left({M\over v}\right)^4
{{v^2}\over{32 \pi^3 m(E-m)\Omega_p (2\Omega_p-E)}}
\label{f}
\EE
\BE
\rho(E)={{(2 m_0^2+m^2)^2}\over{9 m_0^4}}
\left[1-{{2m(E-m)(m_0^2-m^2)}\over{(2m_0^2+m^2)^2}}\right]
\label{rho}
\EE
where $e_\mu (\pp\sigma)$ are the polarization vectors,
$\Omega_p=\sqrt{\pp^2+M^2}$ and $E=2\Omega_{p_0}$.
The right hand side of Eq.(\ref{scattered}) may be interpreted as
the sum of a free wave and a scattered wave for the process
$Z^0Z^0\to h\to Z^0Z^0$. The scattered wave yields\cite{DDe,DDg} 
the cross-section
and the decay width of the Higgs boson which appears as a resonance
for $m>2M$.
Eq.(\ref{scattered}) is an integral equation since, according
to Eq.(\ref{delta}), $\Delta(E)$ is an integral functional of the
wave function $C$. Even in the strong coupling limit, the
small parameter $M/v$ in Eq.(\ref{f}) allows the usual perturbative
expansion obtained by iteration. Thus, at leading order, substituting
Eq.(\ref{scattered}) in Eq.(\ref{delta}) gives
\BE
\Delta(E)={2\over E}
\left(3-{{E^2}\over{M^2}}+{{E^4}\over{4M^4}}\right)+{\cal{O}}(M^4/v^4)
\label{delta2}
\EE

Let us explore this result in the two opposite limits of very weak
($m_0\approx m$) and very strong ($m_0\gg m$) self-coupling.
For $m=m_0$ the coefficient $\rho(E)=1$ and the scattered wave
$\Delta(E)f(\pp,E)$ becomes identical to that obtained 
by Di Leo and Darewych\cite{DDg}. The cross-section is highly
resonant near $E=m$ and can be fitted by the Breit-Wigner
formula yielding\cite{DDg} a decay width $\Gamma_{BW}$ identical to
that obtained from covariant perturbation 
theory\cite{5inDDg,6inDDg,7inDDg}:
\BE
\Gamma_{BW}={{m^3}\over{32\pi v^2}}\left(1+{\cal{O}}(M^2/m^2)\right)
\label{BW}
\EE
Thus in the perturbative limit the present variational calculation
and standard covariant perturbation theory are in perfect agreement.
In the opposite strong-coupling regime we already know that
according to Fig.1 the physical
Higgs mass $m$ can be considerably less than the bare mass $m_0$.
The self-coupling $\lambda$ enter the
scattered wave in Eq.(\ref{scattered})
only through the bare mass $m_0$ in the factor
$\rho(E)$. Even in the very strong coupling limit, $\rho(E)$ does
not change too much and is of order unity. For $m,E\ll m_0$
it takes the limit value $\>\rho(E)\approx 4/9$. 
The non-perturbative decay width
follows\cite{DDg} as
$\Gamma_{NP}=\rho(m)\Gamma_{BW}\approx (4/9)\Gamma_{BW}$.
Thus, apart from the prefactor $\rho$, the decay width is obtained
by inserting the renormalized Higgs mass $m$ in the
standard perturbative result (\ref{BW}). As a consequence,
whatever is the strength of the self-coupling $\lambda$,
if the physical Higgs mass is small the decay width remains
small in the $Z^0-Z^0$ resonance. We do not see how this scenario
could be changed by the inclusion of other processes.

Our findings are not in disagreement with the so called
{\it equivalence theorem}\cite{equiv1,5inDDg,equiv2} 
which states that
at high energies the scattering amplitudes of longitudinal bosons
are equivalent to the scattering amplitudes of their
corresponding would-be Goldstone bosons. In fact, the Higgs boson
and the Goldstone bosons are coupled by the same interaction 
strength $\lambda$, which is assumed to be large in the 
strong-coupling limit. However in the Higgs mechanism 
the Goldstone bosons are not physical since the corresponding
degrees of freedom are taken by the longitudinal polarizations
of the massive vector bosons. It is only at high energy that 
the scattering amplitudes of the longitudinal gauge bosons
are well described by the unphysical amplitudes of the Goldstone
bosons. In the strong-coupling limit the Higgs mass is kept small
by the renormalization effect, and the Higgs resonance at $E=m$
is a low energy process which cannot be described by use of the
equivalence theorem. 

We must mention that, by $1/N$ expansion in the strong-coupling
limit, Ghinculov and Binoth\cite{1/N} find a 
large decay width that
increases with $\lambda$ even beyond the saturation of $m$.
These authors do not explore the very strong coupling
regime where the Higgs mass is small. Besides, their expansion
starts from a $O(N)$ symmetric sigma model which contains the
unphysical Goldstone bosons, and their calculation contains
a tachyonic pole which is regularized by a perturbative method.
Thus it is not clear if their method can be regarded as a genuine
non-perturbative approximation, and if their finding can be compared
to our low energy calculation for the decay width.

The existence of a quite extended strong-coupling range, where the physical
Higgs mass $m$ is small, increases the chances of detecting the Higgs boson
below the TeV scale. However a strongly interacting light Higgs would 
differ from a weakly coupled one for several detectable aspects. For instance
the existence of bound states would be the signature of a strongly
interacting Higgs sector. While parturbation theory would be enough for
a weakly interacting boson, the role of non-perturbative calculations would
be determinant if the Higgs field turns out to be strongly self-coupled.

In summary, by a non-perturbative variational method we have shown that
in the strong-coupling limit the mass of the Higgs boson would be small
as a consequence of mass renormalization. Moreover the decay process
at $E\approx m$ would be a low energy process characterized by a
small decay width. Thus, in order to establish if the Higgs sector is
weakly or strongly interacting, the eventual detection of a light Higgs boson
will not be enough, and the more general phenomenology has to be
considered and compared with the predictions of non-perturbative calculations.

I acknowledge useful conversations with P. Stevenson, M. Consoli,
G. Rupp, D. Zappal\`a, A. Ghinculov and T. Binoth.

\begin{figure}
\caption{The physical Higgs mass $m$ versus the bare mass $m_0$ (which
fixes the coupling strength), for an energy cut-off 
$\Lambda=14$ TeV. The dotted line represents the tree-level 
perturbative approximation $m=m_0$, which only holds in the 
weak-coupling regime $m_0<0.3$ TeV.}
\end{figure}

\begin{figure}
\caption{Higgs-Higgs binding energy $E-2m$ in units of $2m$ 
versus physical Higgs mass $m$ 
for a cut-off $\Lambda=14$ TeV,  according to Eq.(\ref{eqint})
of the text (squares). For comparison, the binding energy obtained 
by the simpler $\vert hh\rangle+\vert hhh\rangle$ trial state is
reported (circles). Notice that the binding energy decreases as the
physical mass increases, since this last one is a decresing function
of the coupling strength according to Fig.1}
\end{figure}

\begin{thebibliography} {99}
\bibitem{perturb} As already discussed in Ref.\cite{varhiggs}, 
  the results of Ref.\cite{DDg} and \cite{DDh} can be recovered
  by a perturbative approximation of our exact eigenvalue equations.
\end{thebibliography}
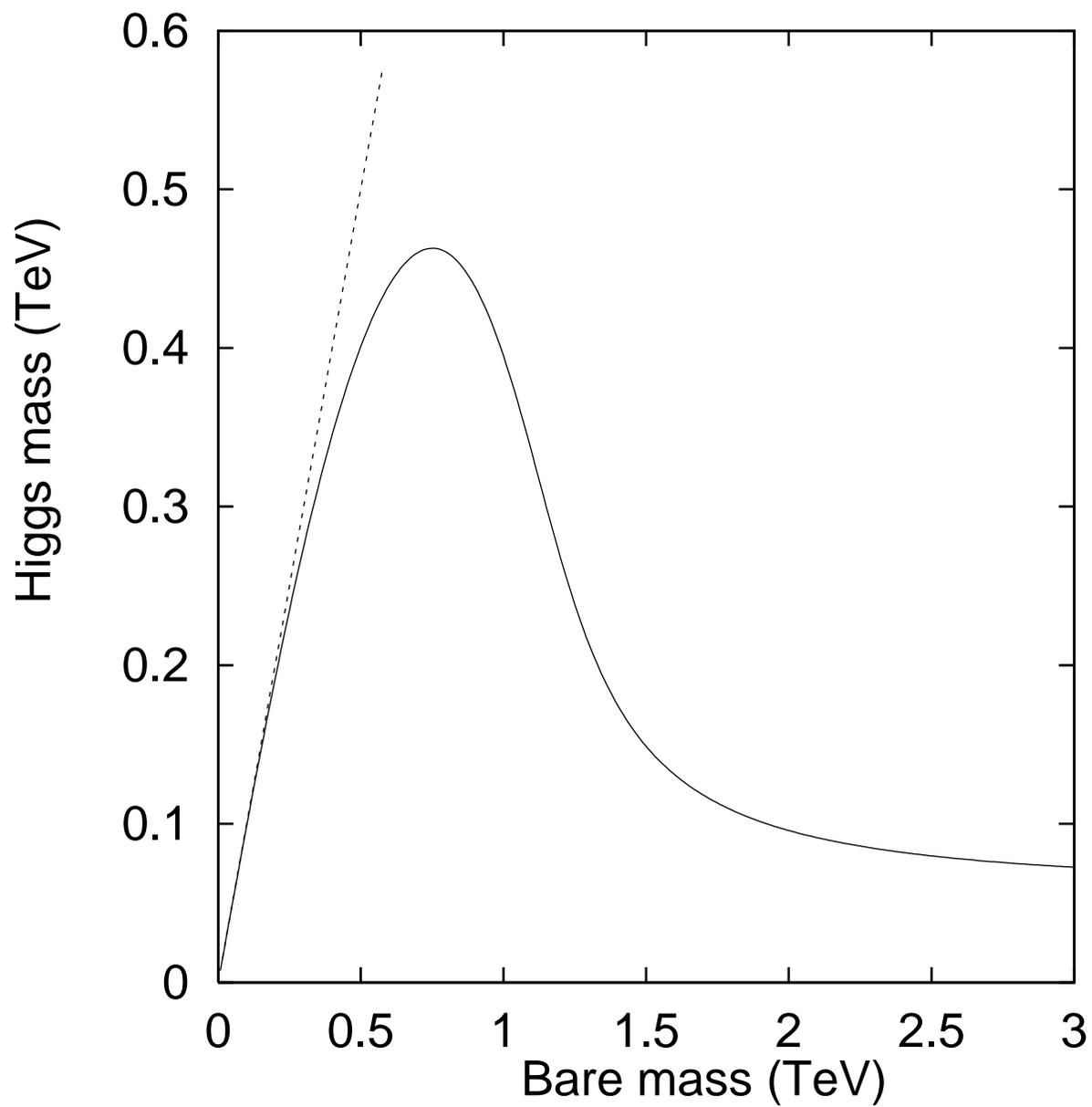
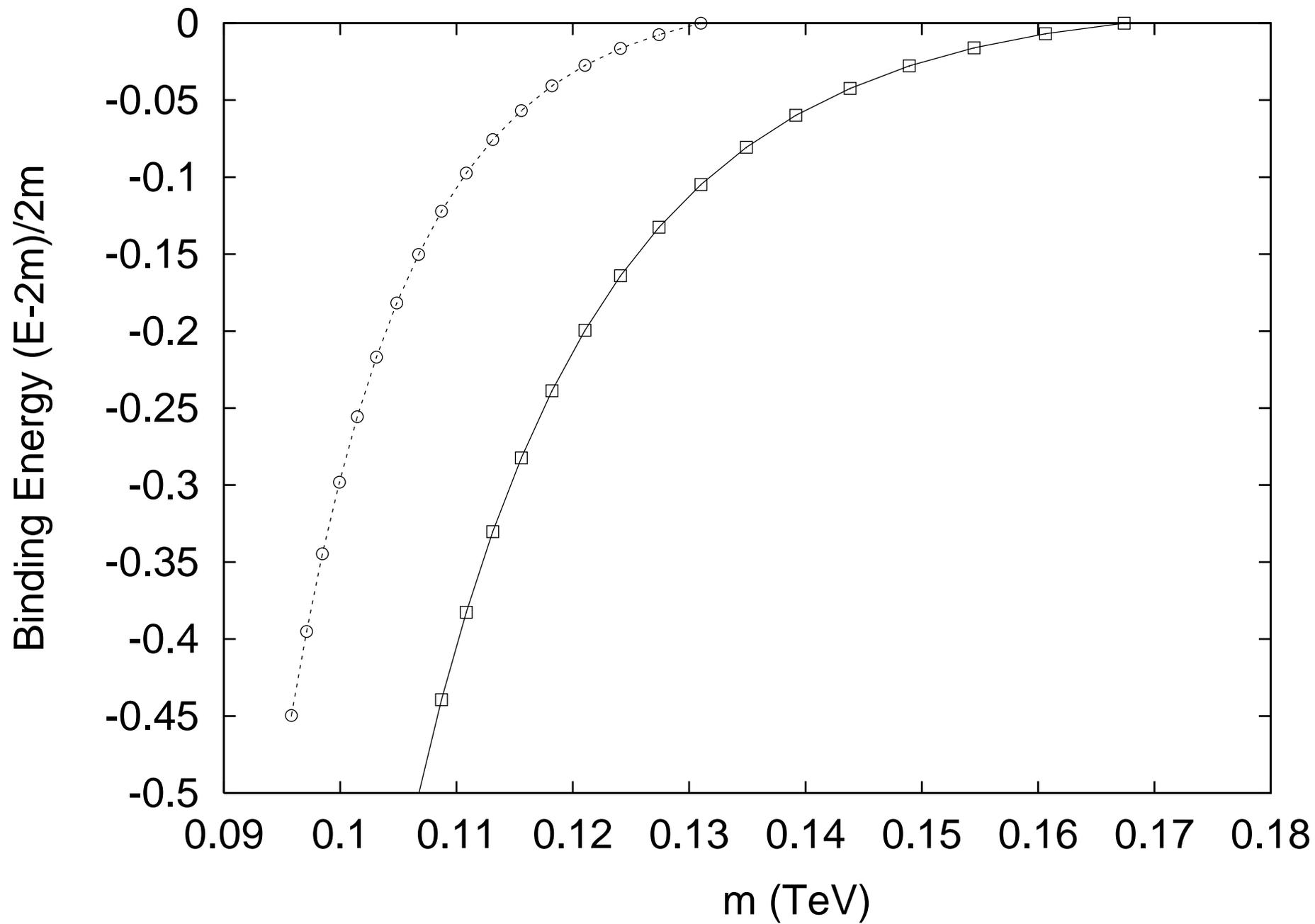
\end{document}